\documentclass[%
 preprint,
 amsmath,amssymb,
 aip,
]{revtex4-1}

\usepackage{graphicx}
\usepackage{dcolumn}
\usepackage{bm}
\usepackage{siunitx}
\usepackage{multirow}
\usepackage{mhchem}

\newcommand{\concat}{\oplus}

\begin{document}


\title{Hierarchical Visualization of Materials Space with Graph Convolutional Neural Networks}

\author{Tian Xie}
\author{Jeffrey C. Grossman}
 \affiliation{Department of Materials Science and Engineering, Massachusetts Institute of Technology, Cambridge, Massachusetts 02139, United States}

\date{\today}

\begin{abstract}

The combination of high throughput computation and machine learning has led to a new paradigm in materials design by allowing for the direct screening of vast portions of structural, chemical, and property space. The use of these powerful techniques leads to the generation of enormous amounts of data, which in turn calls for new techniques to efficiently explore and visualize the materials space to help identify underlying patterns. In this work, we develop a unified framework to hierarchically visualize the compositional and structural similarities between materials in an arbitrary material space with representations learned from different layers of graph convolutional neural networks. We demonstrate the potential for such a visualization approach by showing that patterns emerge automatically that reflect similarities at different scales in three representative classes of materials: perovskites, elemental boron, and general inorganic crystals, covering material spaces of different compositions, structures, and both. For perovskites, elemental similarities are learned that reflects multiple aspects of atom properties. For elemental boron, structural motifs emerge automatically showing characteristic boron local environments. For inorganic crystals, the similarity and stability of local coordination environments are shown combining different center and neighbor atoms. The method could help transition to a data-centered exploration of materials space in automated materials design.
\end{abstract}

\maketitle


\section{Introduction}

Efficient exploration of the materials space has been central to material discovery as a result of the limited experimental and computational resources compared with its vast size. Often compositional or structural patterns are sought from past experiences that might guide the design of new materials, improving the efficiency of material exploration~\cite{niu2015review, snaith2013perovskites, xu2013graphene, butler2013progress, madelung1964physics}. Emerging high-throughput computation and machine learning techniques directly screen large amounts of candidate materials for specific applications~\cite{greeley2006computational, senkan1998high, potyrailo2011combinatorial, curtarolo2013high, hautier2010finding, gomez2016design, faber2016machine, rupp2012fast}, which enables fast and direct exploration of the material space. However, the large quantities of material data generated makes the discovery of patterns challenging with traditional, human-centered approaches. Instead, an automated, data-centered method to visualize and understand a given materials design phase space is needed in order to improve the efficiency of exploration. 

The key in visualizing material space is to map materials with different compositions and structures into a lower dimensional manifold where the similarity between materials can be measured by their Euclidean distances. One major challenge in finding such manifolds is to develop a unified representation for different materials. A widely-used method is representing materials with feature vectors, where a set of descriptors are selected to represent each material~\cite{pilania2013accelerating, meredig2014combinatorial, ward2016general}. There are also methods that automatically select descriptors that are best for predicting a desired target property~\cite{ghiringhelli2015big}. Recent work has also developed atomic-scale representations to map complex atom configurations into low dimensional manifolds, such as atom centered symmetry functions~\cite{behler2011atom}, social permutation invariant (SPRINT) coordinates~\cite{pietrucci2011graph}, global minimum of root-mean-square distance~\cite{sadeghi2013metrics}, smooth overlap of atomic positions (SOAP)~\cite{bartok2013representing}, and many other methods~\cite{schutt2014represent, faber2018alchemical, glielmo2018efficient}. These representations often have physically meaningful parameters that can highlight some structural or chemical features. Often material descriptors and atomic representations are used together to combine compositional and structural information~\cite{ward2017including, faber2018alchemical}. They have been used to visualize the material and molecular similarities\cite{isayev2015materials, de2016comparing, musil2018machine}, as well as explore the complex configurational space of biological systems~\cite{das2006low, ceriotti2011simplifying, spiwok2011metadynamics, rohrdanz2013discovering} and water structures~\cite{pietrucci2015systematic, engel2018mapping}. In addition to Euclidean distances, similarity kernels are also used to compare material similarities~\cite{de2016comparing, musil2018machine}. Combined with machine learning algorithms, these representations were also used to predict material properties~\cite{ghiringhelli2015big, pilania2013accelerating, meredig2014combinatorial, ward2016general, faber2016machine, seko2015prediction, isayev2017universal, schutt2014represent} and construct force fields~\cite{behler2007generalized, bartok2013representing, botu2016machine}.

In parallel to these efforts, the success of ``deep learning'' has inspired a group of representations purely based on neural networks. Instead of designing descriptors or atomic representations that are fixed or contain several physically meaningful parameters, they use relatively general neural network architectures with a large number of trainable weights to learn a representation directly. This field started with building neural networks on molecular graphs~\cite{duvenaud2015convolutional, kearnes2016molecular, gilmer2017neural, schutt2017quantum}, and was recently expanded to periodic material systems by us~\cite{xie2018crystal} and Schutt et al.~\cite{schutt2018schnet}. It has been shown that given large amounts of data, these methods can outperform many other representations on the task of predicting molecular properties~\cite{wu2018moleculenet}. However, the general neural network architecture may also limit performance when the data size is small since there is no material specific information built-in. It is worth noting that many machine learning force fields combine atomic representations and neural networks~\cite{blank1995neural, behler2007generalized, behler2011atom}, but they usually deal with different compositions separately and use a significantly smaller number of weights. It has been shown that the hidden layers of these neural networks can learn physically meaningful representations by proper design of the network architecture. For instance, several works have investigated the ideas of learning atom energies~\cite{xie2018crystal, deringer2018data, schutt2017quantum} and elemental similarities~\cite{schmidt2017predicting, zhou2018learning}. In addition, recent work showed that element similarities can also be learned using a specially designed SOAP kernel~\cite{willatt2018data}. 


In this work, we aim to develop a unified framework to hierarchically visualize the compositional and structural similarities between materials in an arbitrary material space with representations learned from different layers of the neural networks. The network is based on a variant of our previously developed crystal graph convolutional neural networks (CGCNN) framework~\cite{xie2018crystal}, but it is designed to focus on presenting the similarities between materials at different scales, including elemental similarities, local environment similarities, and local energies. We apply this approach to visualize three material spaces: perovskites, elemental boron, and general inorganic crystals, covering material spaces of different compositions, different structures, and both, respectively. We show that in all three cases pattern emerges automatically that might aid in the design of new materials. 

\section{Methods}

\begin{figure}[tbh]
  \centering
  \includegraphics[width=\linewidth]{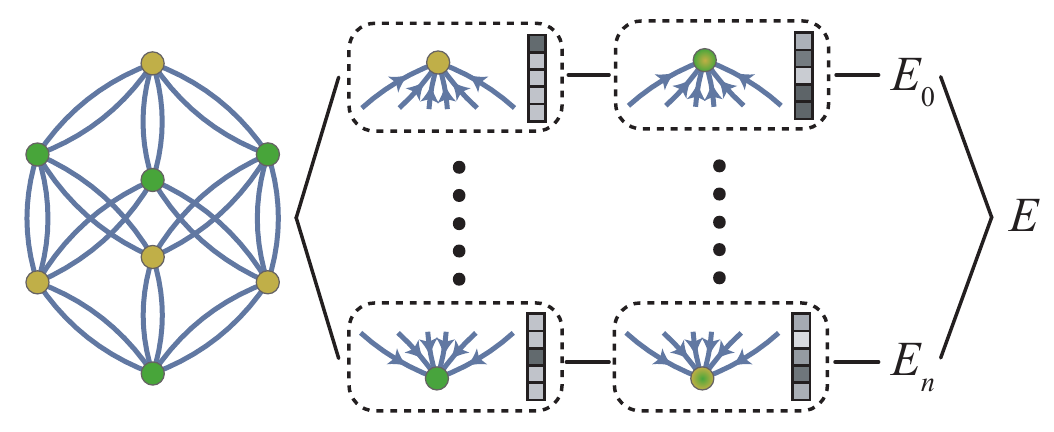}
  \caption{The structure of the crystal graph convolutional neural networks.}
  \label{fig:illustrative}
\end{figure}

To visualize the crystal space at different scales, we design a variant of CGCNN~\cite{xie2018crystal} that has meaningful interpretation at different layers of the neural network. The learned CGCNN network provides a vector representation of the local environments in each crystal that only depends on its composition and structure without any human designed features, enabling us to explore the materials space hierarchically. 

We first represent the crystal structure with a multigraph $\mathcal{G}$ that encodes the connectivity of atoms in the crystal. Each atom is represented by a node $i$ in $\mathcal{G}$ which stores a vector $\bm{v}_i$ corresponding to the element type of the atom. To avoid introducing any human bias, we set $\bm{v}_i$ to be a random 64 dimensional vector for each element and allow it to evolve during the training process. Then, we search for the 12 nearest neighbors for each atom and introduce an edge $(i, j)_k$ between the center node $i$ and neighbor $j$. The subscript $k$ indicates that there can be multiple edges between the same end nodes as a result of the periodicity of the crystal. The edge $(i, j)_k$ stores a vector $\bm{u}_{(i,j)_k}$ whose $t$th element depends on the distance between $i$ and $j$ by,
\begin{equation}
	\bm{u}_{(i,j)_k}[t] = \exp(-(d_{(i,j)_k} - \mu_t)^2 / \sigma^2)
\end{equation}
where $\mu_t = t \cdot \SI{0.2}{\angstrom}$ for $t = 0, 1, ..., 40$ and $\sigma = \SI{0.2}{\angstrom}$.

In graph $\mathcal{G}$, each atom $i$ is initialized by a vector $\bm{v}_i$ whose value solely depends on the element type of atom $i$. We call this iteration 0 where
\begin{equation}
	\bm{v}_i^{(0)} = \bm{v}_i
\end{equation}

Then, we perform convolution operations on the multigraph $\mathcal{G}$ with the convolution function designed in Ref.~\cite{xie2018crystal} which allows atom $i$ to interact with its neighbors iteratively. In iteration $t$, we first concatenate neighbor vectors $\bm{z}_{(i,j)_k}^{(t-1)} = \bm{v}_i^{(t-1)} \concat \bm{v}_j^{(t-1)} \concat \bm{u}_{(i, j)_k}$, and then perform the convolution by,
\begin{multline} \label{eq:conv_fun2}
	\bm{v}_i^{(t)} = \bm{v}_i^{(t-1)} + \sum_{j, k} \sigma(\bm{z}_{(i,j)_k}^{(t-1)} \bm{W}_f^{(t-1)} + \bm{b}_f^{(t-1)}) \\ \odot g(\bm{z}_{(i,j)_k}^{(t-1)} \bm{W}_s^{(t-1)} + \bm{b}_s^{(t-1)}) 
\end{multline}
where $\odot$ denotes element-wise multiplication, $\sigma$ denotes a sigmoid function, and $g$ denotes any non-linear activation function, and $\bm{W}$ and $\bm{b}$ denotes weights and biases in the neural network, respectively. During these convolution operations, $\bm{v}_i^{(t)}$ forms a series of representations of the local environments of atom $i$ at different scales. 

After $K$ iterations, we perform a linear transformation to map $\bm{v}_i^{(K)}$ to a scalar $E_i$,
\begin{equation}
	E_i = \bm{v}_i^{(K)} \bm{W}_l + b_l
\end{equation}
and then use a normalized sum pooling to predict the averaged total energy per atom of the crystal,
\begin{equation} \label{eq:pool}
	E = \frac{1}{n} \sum_i E_i
\end{equation}
where $n$ is the number of atoms in the crystal. This introduces a physically meaningful term $E_i$ to represent the energy of the local chemical environment. 

The model is trained by minimizing the squared error between predicted properties relative to the DFT calculated properties using backpropagation and stochastic gradient descent. 

In this CGCNN model, each vector represents the local environment of each atom at different scales. Here, we focus three vectors that has the most representative physical interpretations.
\begin{enumerate}
	\item \textit{Element representation} $\bm{v}_i^{(0)}$ that depends completely on the type of element that atom $i$ is composed of, describing the similarities between elements.
	\item \textit{Local environment representation} $\bm{v}_i^{(K)}$ that depends on atom $i$ and its $K$th order neighbors, describing the similarities between local environments that combines the compositional and structural information.
	\item \textit{Local energy representation} $E_i$ that describes the energy of atom $i$. 
\end{enumerate}

\section{Results and Discussions}

To illustrate how this method can help visualize the compositional the structural aspects of the crystal space, we apply it to three datasets that representing different material spaces. 1) a group of perovskite crystals that share the same structure type but have different compositions; 2) different configurations of elemental boron that share the same composition but have different structures; and 3) inorganic crystals from the Materials Project~\cite{Jain2013} that have both different compositions and different structures. 

\begin{figure}[tbh]
  \centering
  \includegraphics[width=\linewidth]{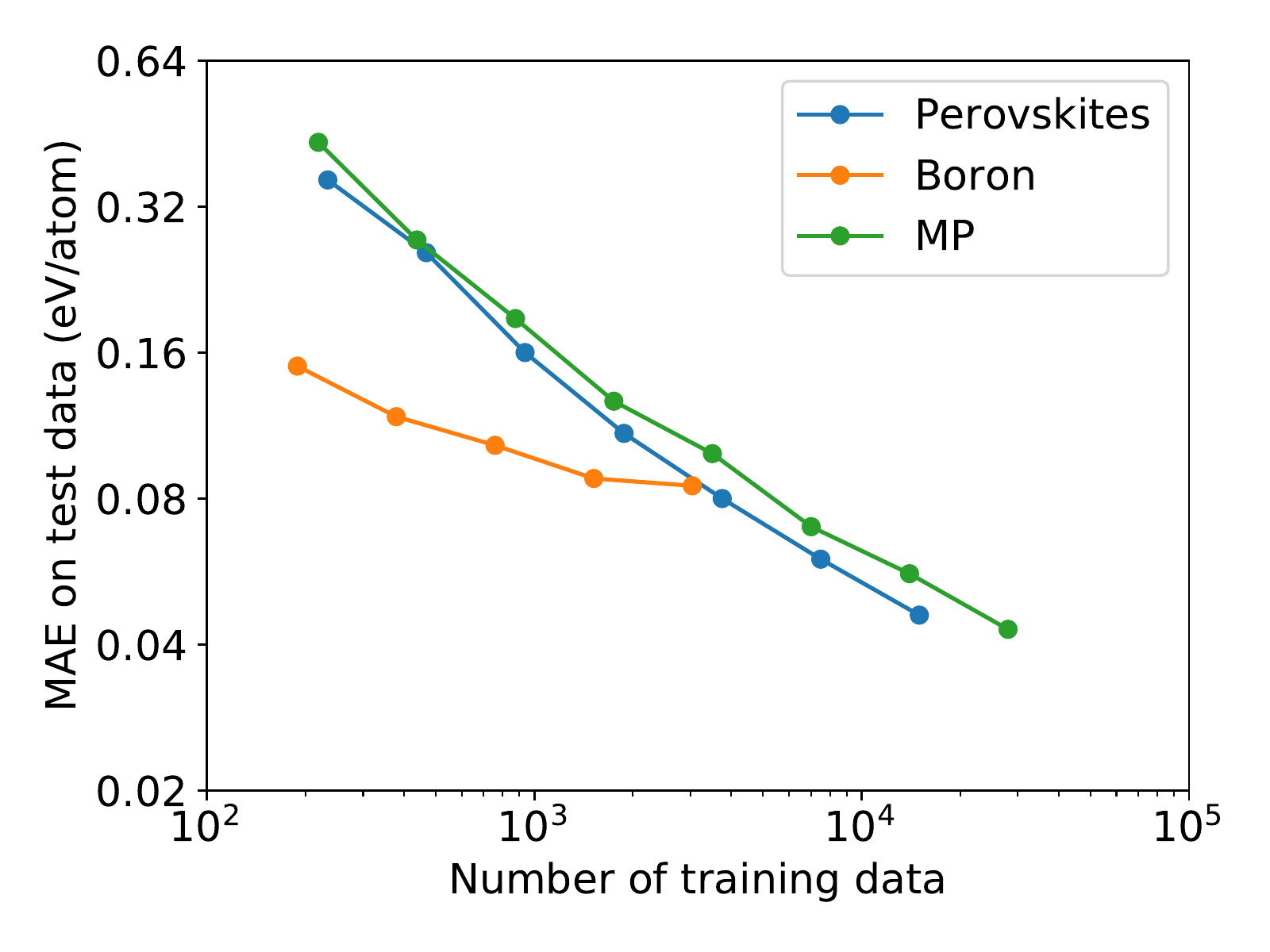}
  \caption{Learning curves for the three representative material spaces. The mean absolute errors (MAEs) on test data is shown as a function of the number of training data for the perovskites~\cite{castelli2012new, castelli2012computational}, elemental boron~\cite{deringer2018data}, and materials project~\cite{Jain2013} datasets.}
  \label{fig:learning-curves}
\end{figure}

For each material space, we train the CGCNN model with 60\% of the data to predict the energy per atom of the materials. 20\% of the data are used to select hyperparameters of the model and the last 20\% are reserved for testing. In Fig.~\ref{fig:learning-curves}, we show the learning curves for the three representative material spaces where a subset of training data is used to show how the number of training data affects the model prediction performance. As we will show below, the representations learned by predicting the energies automatically gain physical meanings and can be used to explore the materials spaces.

\subsection{Perovskite: compositional space}

\begin{figure*}[tbh]
  \centering
  \includegraphics[width=\linewidth]{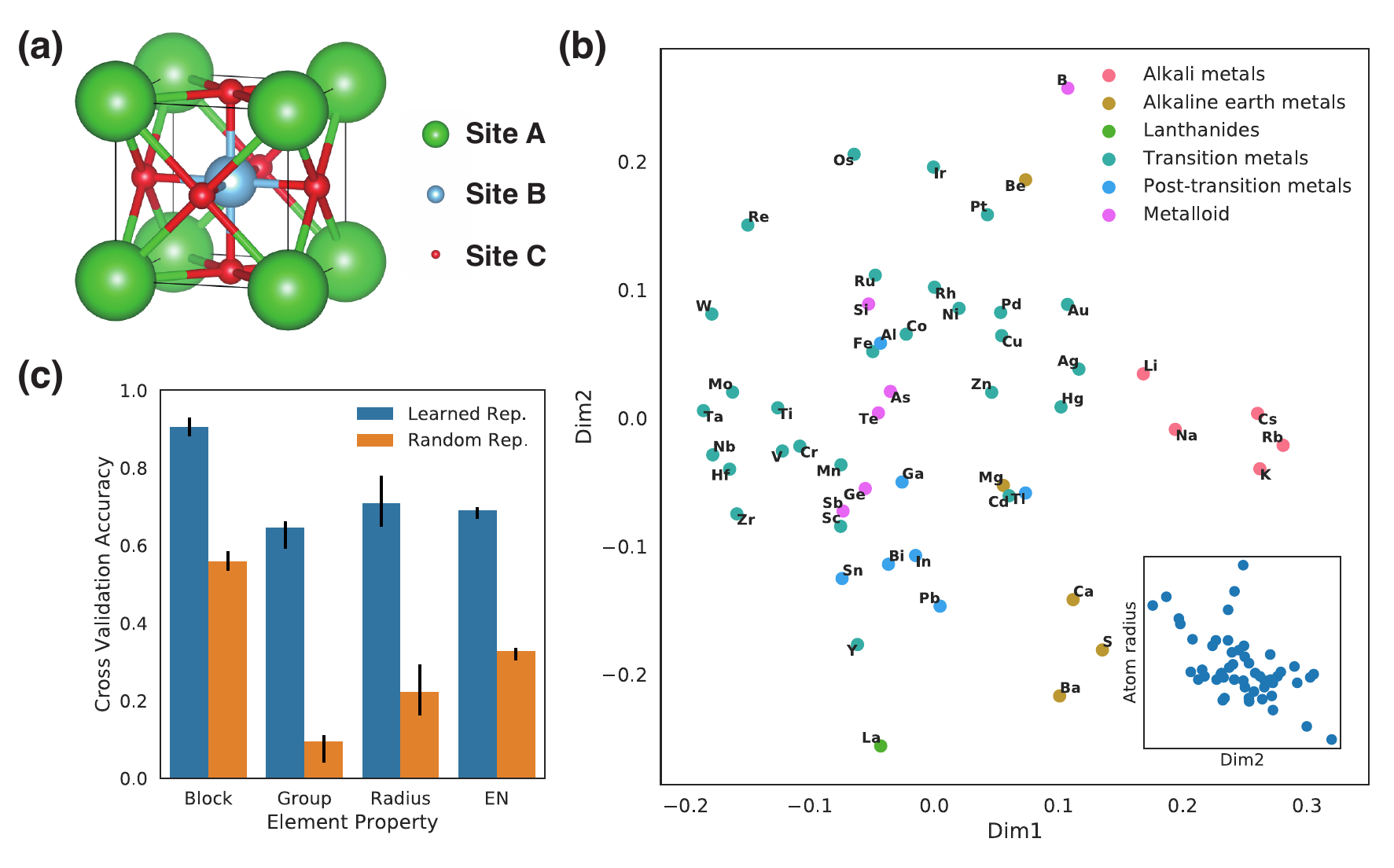}
  \caption{Visualization of the element representations learned from the perovskite dataset. (a) The perovskite structure type. (b) Visualization of the two principal dimensions with principal component analysis. (c) Prediction performance of several atom properties using a linear model on the element representations. }
  \label{fig:perovskite}
\end{figure*}

First, we explore the compositional space of perovskites by visualizing the \textit{element representations}. Perovskite is a crystal structure type with the form of \ce{ABC_3} as shown in Fig.~\ref{fig:perovskite}(a). The dataset~\cite{castelli2012new, castelli2012computational} that we used includes 18,928 different perovskites where the elements \ce{A} and \ce{B} can be any nonradioactive metals and the element \ce{C} can be one or several from \ce{O}, \ce{N}, \ce{S}, and \ce{F}. We trained our model to predict the energy above hull with 15,000 training data, and after hyperparameter optimization on 1,890 validation data, we achieve a prediction mean absolute error (MAE) of 0.042 eV/atom on 2,000 test data. The prediction performance is excellent and lower than several recent ML models such as those of Schmidt et al. (0.121 eV/atom)~\cite{schmidt2017predicting} and Xie et al. (0.099 eV/atom)~\cite{xie2018crystal}. The learning curve in Fig. \ref{fig:learning-curves} shows a straight line in log-log scale, indicating a steady increase of prediction performance as the number of training data increases. 

In Fig. \ref{fig:perovskite}(b)(c), the element representation $\bm{v}_i^{(0)}$, a 64 dimensional vector, is visualized for every nonradioactive metal element after training with the perovskite dataset. Fig. \ref{fig:perovskite}(b) shows the projection of these element representations on a 2D plane using principal component analysis, where elements are colored according to their elemental groups. We can clearly see that similar elements are grouped together based on their stability in perovskite structures. For instance, alkali metals are grouped on the right of the plot due to their similar properties. The large alkaline earth metals (Ba, Se, and Ca) are grouped on the bottom, distinct from Mg and Be, because their larger radius stabilizes them in the perovskite structure. On the left side are elements such as W, Mo, and Ta that favor octahedral coordinations due to their configuration of \textit{d} electrons, which might be related to their extra stability in the B site~\cite{xie2018crystal}. Interestingly, we can also observe a trend of decreasing atom radius from the bottom of the plot to the top as shown in the insert of Fig. \ref{fig:perovskite}(b), except for the alkali metals as outliers. This indicates that CGCNN learns the atom radius as an important feature for perovskite stability. Recently, Schutt et al. also discovered similar grouping of elements with data from the Materials Project~\cite{schutt2018schnet}. In general, these visualizations can help discover similarities between elements for designing novel perovskite structures.

We also study how the element representations evolve as the number of training data changes. In Fig. S1, we show the 2D projections of the element representations when 234, 937, 3,750, and 15,000 training data are used, respectively. The projection looks completely random with 234 training data, and some patterns start to emerge when 937 training data are used. In Fig. S1(b), transition metals are grouped on top of the figure while large metals like La, Ca, Sr, Ba, and Cs are grouped at the bottom. With 3,750 training data, the figure is already close to Fig. \ref{fig:perovskite}(b) and the relation between atom radius and the second dimension is clear. Fig. \ref{fig:perovskite}(b) and Fig. S1(d) are almost identical after rotations because they both use 15,000 training data. Note that these representations start from different random initializations, but they result in similar patterns after training with the same perovskite data. 

However, these 2D plots only account for part of the 64-dimensional element representation vectors. To fully understand how element properties are learned by CGCNN, we use linear logistic regression (LR) models to predict the block type, group number, radius, and electronegativity of each element from their learned representation vectors. In Fig. \ref{fig:perovskite}(c), we show the 3-fold cross validation accuracy of the LR models and compare them with LR models learned from random representations, which helps to rule out the possibility that the predictions are caused by coincidences. We discover a significantly higher prediction accuracy of the learned representations for all four properties, demonstrating that the element representations can reflect multiple aspects of element properties. For instance, the model predicts the block of the element with over 90\% accuracy, and the same representation also predicts the group number, radius, and electronegativity with over 60\% accuracy. This is surprising considering that there are 16 different elemental groups represented. It is worth noting that these representations are learned only from the perovskite structures and the total energy above hull, but they are in agreement with these empirical element properties reflecting decades of human chemical intuition.

\subsection{Elemental boron: structural space}

\begin{figure}[tbh]
  \centering
  \includegraphics[width=\linewidth]{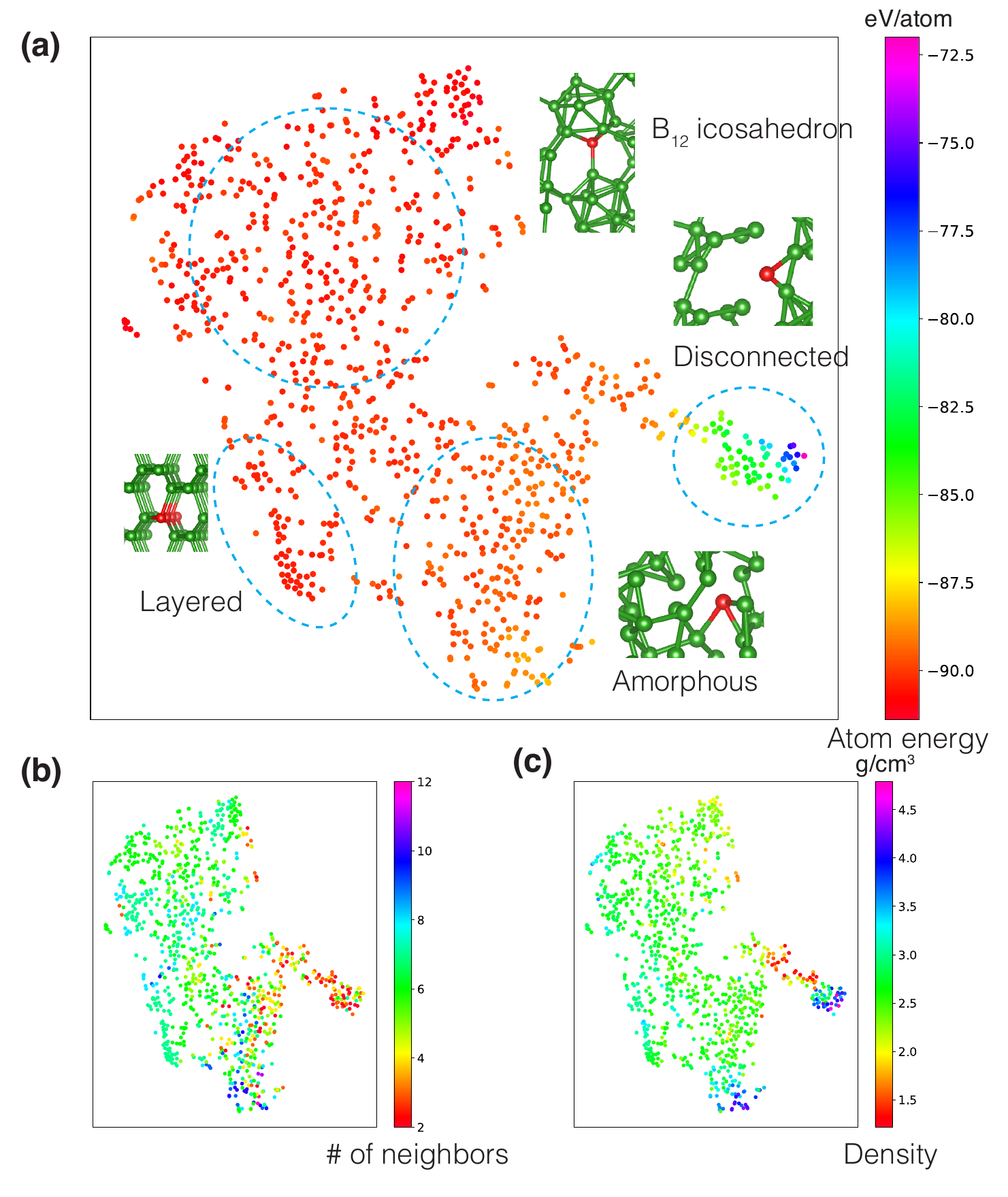}
  \caption{Visualization of the local environment representations learned from the elemental boron dataset. The original 64D vectors are reduced to 2D with the t-distributed stochastic neighbor embedding algorithm. The color of each plot is coded with learned local energy (a), number of neighbors calculated by Pymatgen package~\cite{ONG2013314} (b), and density (c). Representative boron local environments are shown with the center atom colored in red. }
  \label{fig:B}
\end{figure}

As a second example, we explore the structural space of elemental boron by visualizing the \textit{local environment representations} and the corresponding \textit{local energies}. Elemental boron has a number of complex crystal structures due to its unique, electron-deficient bonding nature~\cite{ogitsu2013beta, deringer2018data}. We use a dataset that includes 5038 distinct elemental boron structures and their total energies calculated using density functional theory~\cite{deringer2018data}. We train our CGCNN model with 3038 structures, and perform hyperparameter optimization with 1000 validation structures. The MAE of predicted energy relative to DFT results on the remaining 1000 test structures is 0.085 eV/atom. The learning curve in Fig. \ref{fig:learning-curves} shows a much smaller slope compared with the other material spaces. One explanation is that there exist many highly unstable boron structures in the dataset, whose energies might be hard to predict given the limited structures covered by the training data.

In Fig. \ref{fig:B}, 1000 randomly sampled boron local environment representations are visualized in 2 dimensions using the t-distributed stochastic neighbor embedding (t-SNE) algorithm~\cite{maaten2008visualizing}. We observe primarily four different regions of different boron local environments, and we discover a smooth transition of local energy, number of neighbor atoms, and the density between different regions. The disconnected region consists of boron atoms at the edge of boron clusters [Fig. S1(a-c)]. These atoms have very high local energies and lower number of neighbors, as to be expected, and their density varies depending on the distances between clusters. The amorphous region includes boron atoms in a relatively disordered local configuration, and their local energies are lower than the disconnected counterparts but higher than other other configurations [Fig. S1(d-f)]. We can see that the number of neighbors fluctuates drastically in these two regions due to the relatively disordered local structures. The layered region is composed of boron atoms in layered boron planes, where neighbors on one side are closely bonded and the neighbors on the other side are further away [Fig. S1(g-i)]. The \ce{B_{12}} icosahedron region includes boron local environments with the lowest local energy, which have a characteristic icosahedron structure [Fig. S1(j-l)]. The local environments in each region share common characteristics but are slightly different in detail. For instance, most boron atoms in the \ce{B_{12}} icosahedron region are in a slightly distorted icosahedron, and the local environments in Fig. S1(l) only have certain features of an icosahedron. Note that these representations are rather localized. The global structure of Fig. S1(c) is layered, but the representation of the highlighted atom at the edge is closer to the disconnected region locally. Some experimentally observed boron structures, like boron fullerenes, are not presented in the dataset. We calculate the local environment representations of every distinct boron atom of two boron fullerenes~\cite{zhai2014observation} using the trained CGCNN, and plot them into the original 2D visualization in Fig. S3. They form a small cluster close to the \ce{B_{12}} icosahedron region. This can be explained by the fact that they share many common characteristics to the \ce{B_{12}} icosahedron structure. In addition, the representations of the less symmetric \ce{B_40}($C_s$) are more spread out than the more symmetric \ce{B_40}($D_{2d}$). Note that the pattern in Fig. S3 is slightly different from that in Fig. \ref{fig:B} due to the random nature of the t-SNE algorithm, but the overall structure of the patterns is preserved. 

Taken together, such a visualization approach provides a convenient way to explore complex boron configurations, enabling the identification of characteristic structures and systematic exploration of structural space.

\subsection{Materials Project: compositional and structural space}

\begin{figure*}[tbh]
  \centering
  \includegraphics[width=\linewidth]{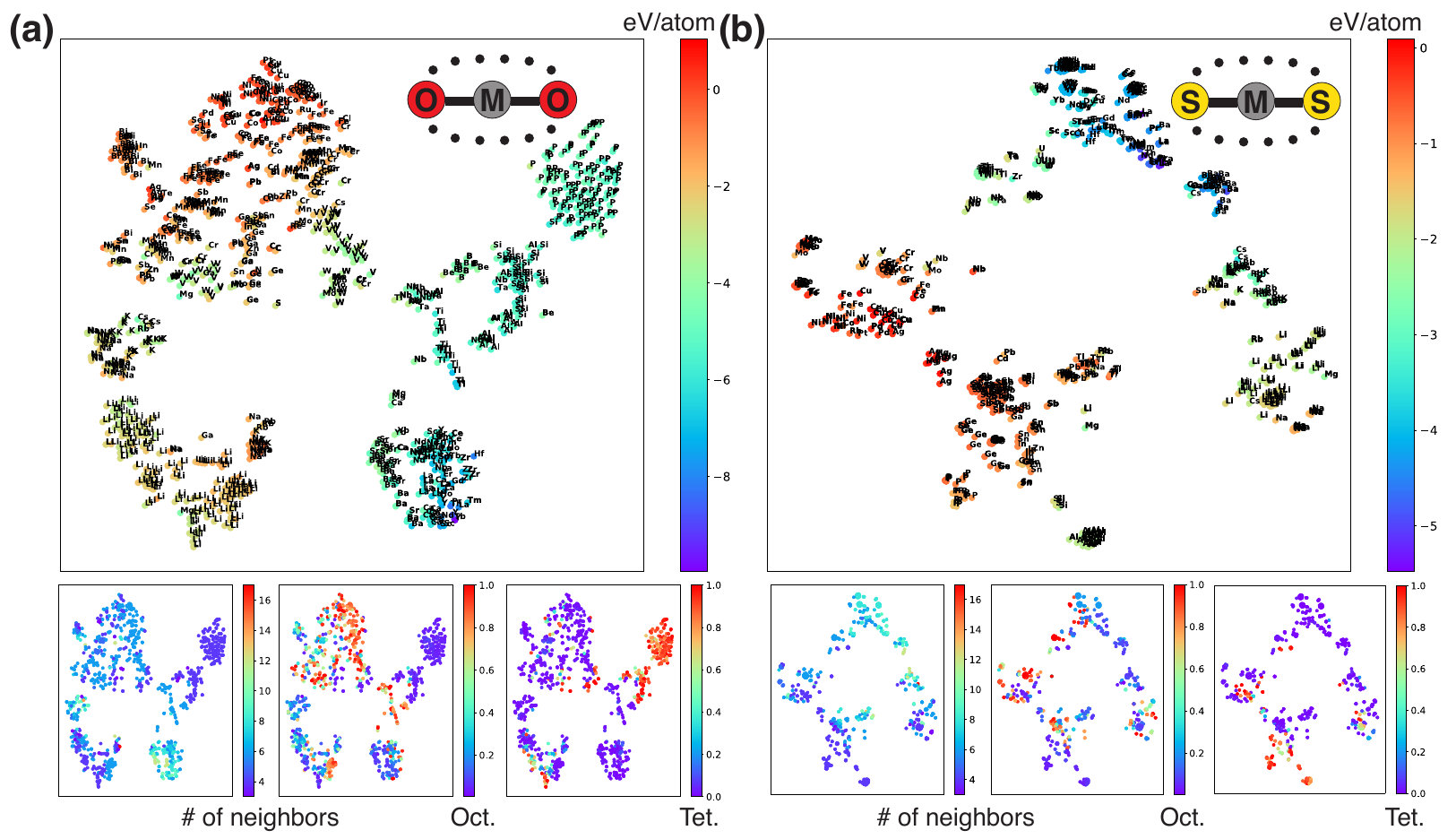}
  \caption{Visualization of the local oxygen (a) and sulfur (b) coordination environments. The points are labelled according to the type of the center atoms in the coordination environments. The colors of the upper parts are coded with learned local energies, and the color of the lower parts are coded with number of neighbors~\cite{ONG2013314}, octahedron order parameter, and tetrahedron order parameter~\cite{zimmermann2017assesing}.}
  \label{fig:chem-env}
\end{figure*}

As a third example of applying this approach, we explore the material space of crystals in the Materials Project dataset~\cite{Jain2013}, which includes both compositional and structural differences, by visualizing the \textit{element representation}, \textit{local environment representation}, and the \textit{local energy representation}. The dataset includes 46744 materials that cover the majority of crystals from the Inorganic Crystal Structure Database~\cite{hellenbrandt2004inorganic}, providing a good representation of known inorganic materials. After training with 28046 crystals and performing hyperparameter optimization with 9348 crystals, our model achieves MAE of predicted energy relative to DFT calculations on the 9348 test crystals of 0.042 eV/atom, slightly higher than the MAE of our previous work, 0.039 eV/atom, with a CGCNN structure focusing on prediction performance~\cite{xie2018crystal}. The learning curve in Fig. \ref{fig:learning-curves} is similar to that of the perovskites dataset, which might indicate a similar prediction performance to the datasets that are composed of stable inorganic compounds. In Table \ref{tab:predictions}, we compare the prediction performance of this method with several recently published works.

\begin{table}
\caption{Comparison of the prediction performance of formation energy per atom. The mean absolute errors (MAEs) on test data reported in several recent works are summarized. Data come from several different but similar inorganic crystal material datasets. MP represents materials project~\cite{Jain2013}, OQMD represents the open quantum materials database~\cite{saal2013materials}, and the ternary compounds are \ce{A_xB_yC_z} compounds calculated by Ref.~\cite{meredig2014combinatorial}.\label{tab:predictions}}
\begin{ruledtabular}
\begin{tabular}{cccc}
  Method    & MAE (eV/atom)		& Data source & Training size\\
  \hline
  This work	& 0.042 	& MP		& 28,046 \\
  CGCNN~\cite{xie2018crystal}  & 0.039 & MP	& 28,046\\
  SchNet~\cite{schutt2018schnet} & 0.035 & MP & 60,000 \\
  Generalized Coulomb matrix~\cite{faber2015crystal}	& 0.37 & MP & 3,000 \\
  Decision trees + heuristic~\cite{meredig2014combinatorial} & 0.12 & Ternary compounds	& 15,000 \\
  Voronoi + composition~\cite{ward2017including} & 0.08 & OQMD & 30,000 \\
  QML~\cite{faber2018alchemical} & $\sim$0.11 & OQMD & 2,000 \\
  Random subspace + REPTree~\cite{ward2016general}	& 0.088 & OQMD & 228,676\\

\end{tabular}
\end{ruledtabular}
\end{table}

In Fig. S2, the element representation of 89 elements learned from the dataset is shown using the same method as that used to generate Fig. \ref{fig:perovskite}(b). We observe similar grouping of elements from the same elemental groups, but the overall pattern differs since it reflect the stability of each element in general inorganic crystals rather than perovskites. For instance, the non-metal and halogen elements stand out because their properties deviates from other metallic elements. 

To illustrate how the compositional and structural spaces can be explored simultaneously, we visualize the oxygen and sulfur coordination environments in the Materials Project dataset using the local environment representation and local energy. 1000 oxygen and 803 sulfur coordination environments are randomly selected and visualized using the t-SNE algorithm. As shown in Fig. \ref{fig:chem-env}(a), the oxygen coordination environments are clustered into 4 major groups. The upper right group has the center atom of non-metal elements like P, Al, Si, forming tetrahedron coordinations. The center atoms of the upper left environments are mostly transition metals, and they mostly form octahedron coordinations. The lower left group has center atoms of alkali metals, and the lower right group has those of alkaline earth metals and lanthanides which have larger radii and therefore higher coordination numbers. The sulfur coordination environment visualization [Fig. \ref{fig:chem-env}(b)] shares similar patterns due to the similarities between oxygen and sulfur, and a similar four-cluster structure can be observed. However, instead of non-metal elements, the lower center group has center atoms of metalloids like Ge, Sn, Sb, since these elements will be more stable in a sulfur vs. oxygen coordination environment.

The local energy of oxygen and sulfur coordination environments are determined by their relative stability to the pure elemental states since the model is trained using the formation energy data, which treats the pure elemental states as the reference energy states. In Fig. S3, we show the change of local energy of oxygen and sulfur local energies as a function of atomic number. We can clearly see that it follows a similar trend as the electronegativity of the elements: elements with lower electronegativity tend to have lower local energy and vice versa. This is because elements with lower electronegativity tends to give the oxygen and sulfur more electrons and thus form stronger bonds. The local energies of alkali metals are slightly higher since they form weaker ionic bonds due to lower charges. Interestingly, the strong covalent bonds between oxygen and Al, Si, P, S forms a V-shaped curve in the figure, with Si-O environments having the lowest energy, contrasting the trend of electronegativity and sulfur coordination environments, whose local energies are dominated by the strength of ionic bonds. We also observe a larger span of local energies in oxygen coordination environments than their sulfur counterparts due to the stronger ionic interactions. 

\begin{figure*}[tbh]
  \centering
  \includegraphics[width=\linewidth]{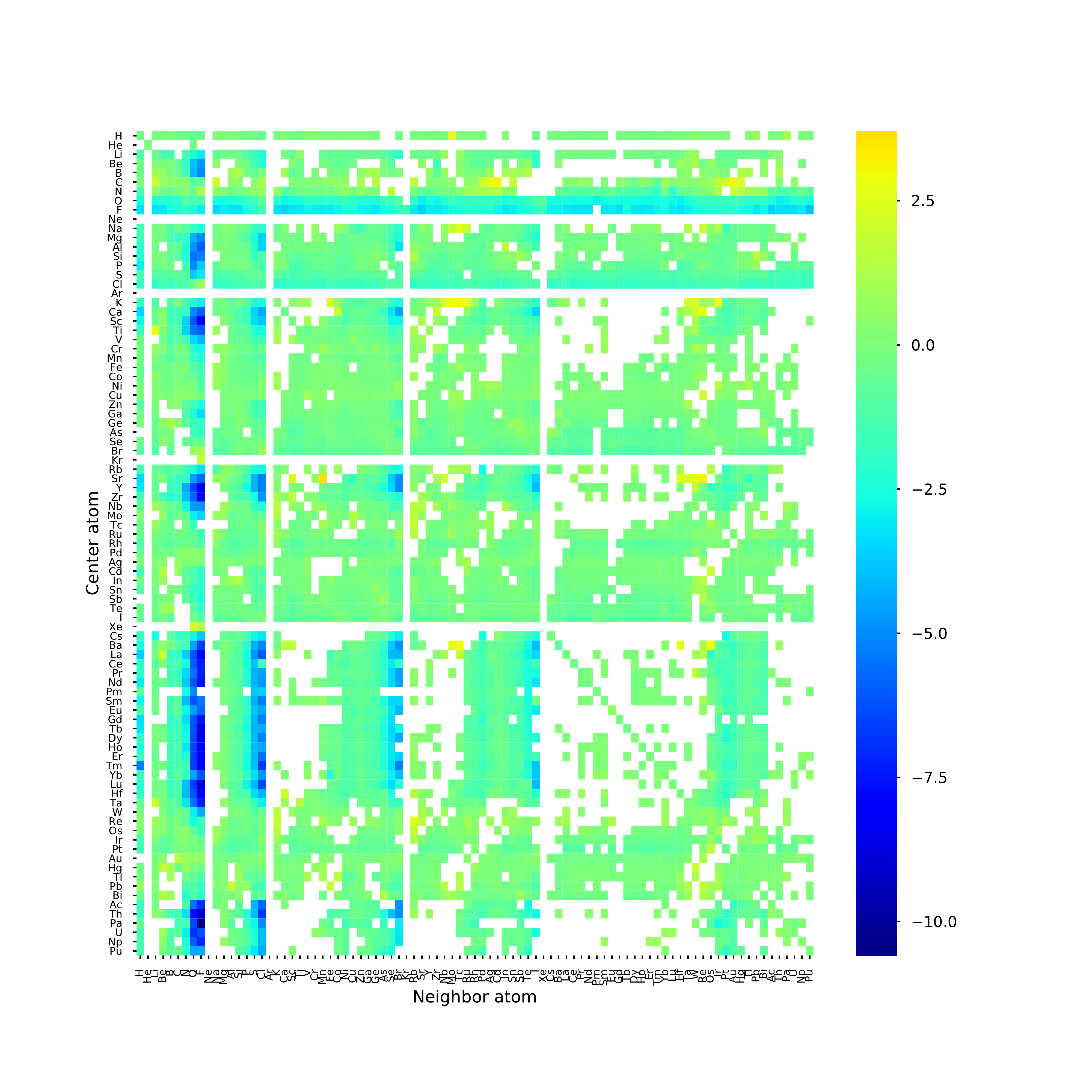}
  \caption{The averaged local energy of 734,077 distinct coordination environments in the Materials Project dataset. The color is coded with the average of learned local energies while having the corresponding elements as the center atom and the first neighbor atom. White is used when no such coordination environment exists in the dataset.}
  \label{fig:chem-env-stability-map}
\end{figure*}

Inspired by these results, we visualize the averaged local energy of 734,077 distinct coordination environments in the Materials Project by combining different center and neighbor atoms in Fig. \ref{fig:chem-env-stability-map}. This figure illustrates the stability of the local coordination environment while combining the corresponding center and neighbor elements. The diagonal line represents coordination environments made up with the same elements with local energy close to zero, which corresponds to elemental substances with zero formation energy. The coordination environments with lowest local energy consist of high valence metals and high electronegativity non-metals, which can be explained by the large cohesive energies due to strong ionic bonds. One abnormality is the stable Al-O, Si-O, P-O, S-O coordination environments, although this can be attributed to their strong covalent bonds. We can also see that Tm-H coordination stands out as a stable hydrogen solid solution~\cite{bonnet1979study}. It is worth noting that each local energy in Fig. \ref{fig:chem-env-stability-map} is the average of many coordination environments with different shape and outer layer chemistry, and we can obtain more information by using additional visualizations similar to Fig. \ref{fig:chem-env}.

\section{Conclusion}

In summary, we developed a unified approach to visualize the compositional and structural space of materials. The method provides hierarchical representations of the local environments at different scales, which enables a general framework to explore different material systems and measure material similarities. The insights gained from the visualizations could help to discover patterns from a large pool of candidate materials that may be impossible by human analysis, and provide guidance to the design of new materials. In addition to energies, this method can potentially be applied to other material properties for the exploration of novel functional materials.

\section{Supplementary Material}

See supplementary material for the details of the hyperparameters for each model, results of the effects of the number of training data on element representations, additional figures showing the structures of boron local environments and the location of boron fullerene local environment representations with respect to the representations of other boron structures, results of the element representations learned from the Materials Project dataset, and results of the change of local energy as a function of atomic number.  

\begin{acknowledgements}
This work was supported by Toyota Research Institute. Computational support was provided through the National Energy Research Scientific Computing Center, a DOE Office of Science User Facility supported by the Office of Science of the U.S. Department of Energy under Contract No. DE-AC02-05CH11231, and the Extreme Science and Engineering Discovery Environment, supported by National Science Foundation grant number ACI-1053575.
\end{acknowledgements}

\bibliography{refs}

\widetext
\clearpage

\begin{center}
\textbf{\large Supplemental Material: Hierarchical Visualization of Materials Space with Graph Convolutional Neural Networks}
\end{center}
\setcounter{equation}{0}
\setcounter{figure}{0}
\setcounter{table}{0}
\setcounter{page}{1}
\setcounter{section}{0}
\makeatletter
\renewcommand{\theequation}{S\arabic{equation}}
\renewcommand{\thefigure}{S\arabic{figure}}
\renewcommand{\bibnumfmt}[1]{[S#1]}
\renewcommand{\citenumfont}[1]{S#1}
\section{Supplementary Methods}

\subsection{Logistic Regression Models}

In the perovskite dataset, we use logistic regression models to predict four different elemental properties. We treat all four predictions as classification problem for consistency, although some of the properties have continuous values. We summarized the categories of each elemental properties in Table~\ref{tab:lg}.

\clearpage
\section{Supplementary Tables}

\begin{table}[tbh]
\caption{Hyperparameters selected for each dataset.\label{tab:hyper}}
\begin{ruledtabular}
\begin{tabular}{llll}
  Dataset    & \# of convolutional layers		& Length of representation $\bm{v}_i^{(t)}$ & learning rate\\
  \hline
  Perovskites	& 4	&	64	& 0.005 \\
  Elemental B	& 4	&	64	& 0.005 \\
  Materials Project	& 4	& 64	& 0.005 \\

\end{tabular}
\end{ruledtabular}
\end{table}

\begin{table}[tbh]
\caption{The categories of each elemental property logistic regression models.\label{tab:lg}}
\begin{ruledtabular}
\begin{tabular}{llp{8cm}}
  Elemental property    & \# of categories		& Categories \\
  \hline
  Block	& 3	& s, p, d\\
  Group	& 16	& 1, 2, ..., 16 \\
  Radius (\SI{}{\angstrom})~\cite{cordero2008covalent}	& 5	& [83, 116), [116, 148), [148, 180), [180, 212), [212, 244) \\
  Electronegativity~\cite{mendeleev2014}	& 5	& [0.788, 1.112), [1.112, 1.434), [1.434, 1.756), [1.756, 2.078), [2.078, 2.4) \\
\end{tabular}
\end{ruledtabular}
\end{table}

\clearpage
\section{Supplementary Figures}

\begin{figure}[tbh]
  \centering
  \includegraphics[width=\linewidth]{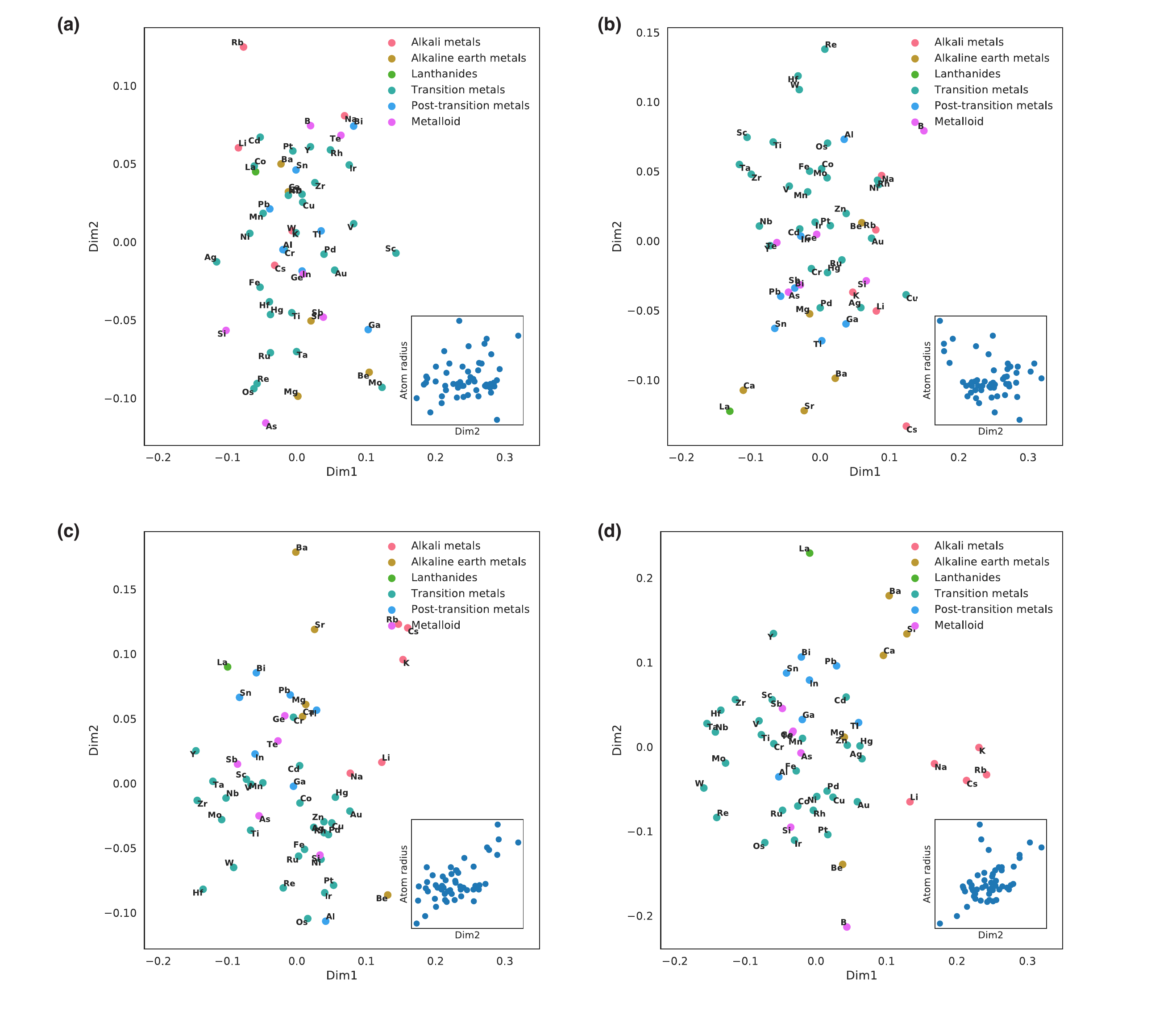}
  \caption{The evolution of element representations as the number of training data increases. The number of training data used are (a) 234, (b) 937, (c) 3,750, (d) 15,000.}
\end{figure}

\begin{figure}[tbh]
  \centering
  \includegraphics[width=\linewidth]{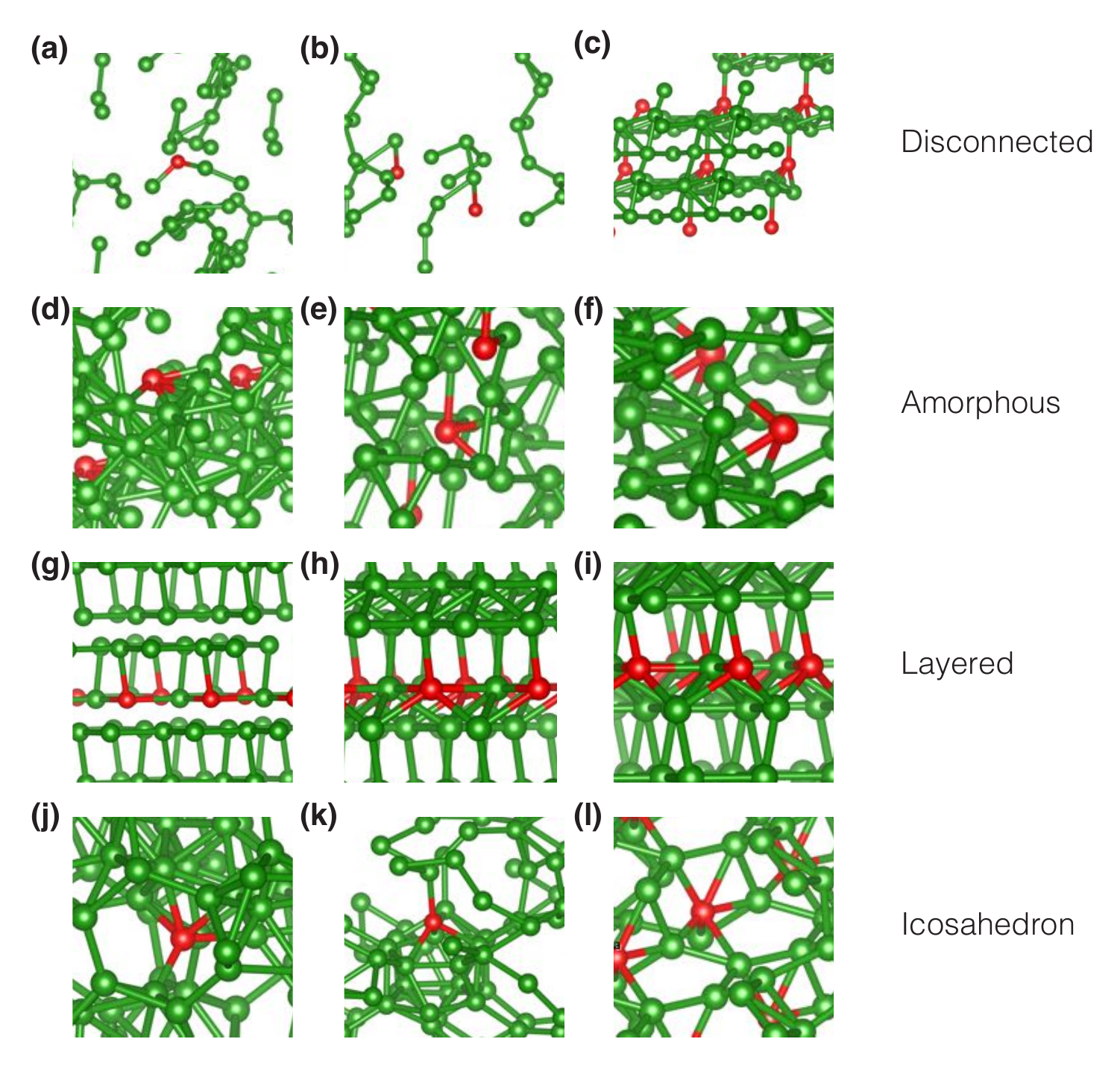}
  \caption{Example local environments of elemental boron in the four regions: (a-c) disconnected, (d-f) amorphous, (h-i) layered, and (j-l) icosahedron. }
  \label{fig:si-B-locals}
\end{figure}

\begin{figure}[tbh]
  \centering
  \includegraphics[width=\linewidth]{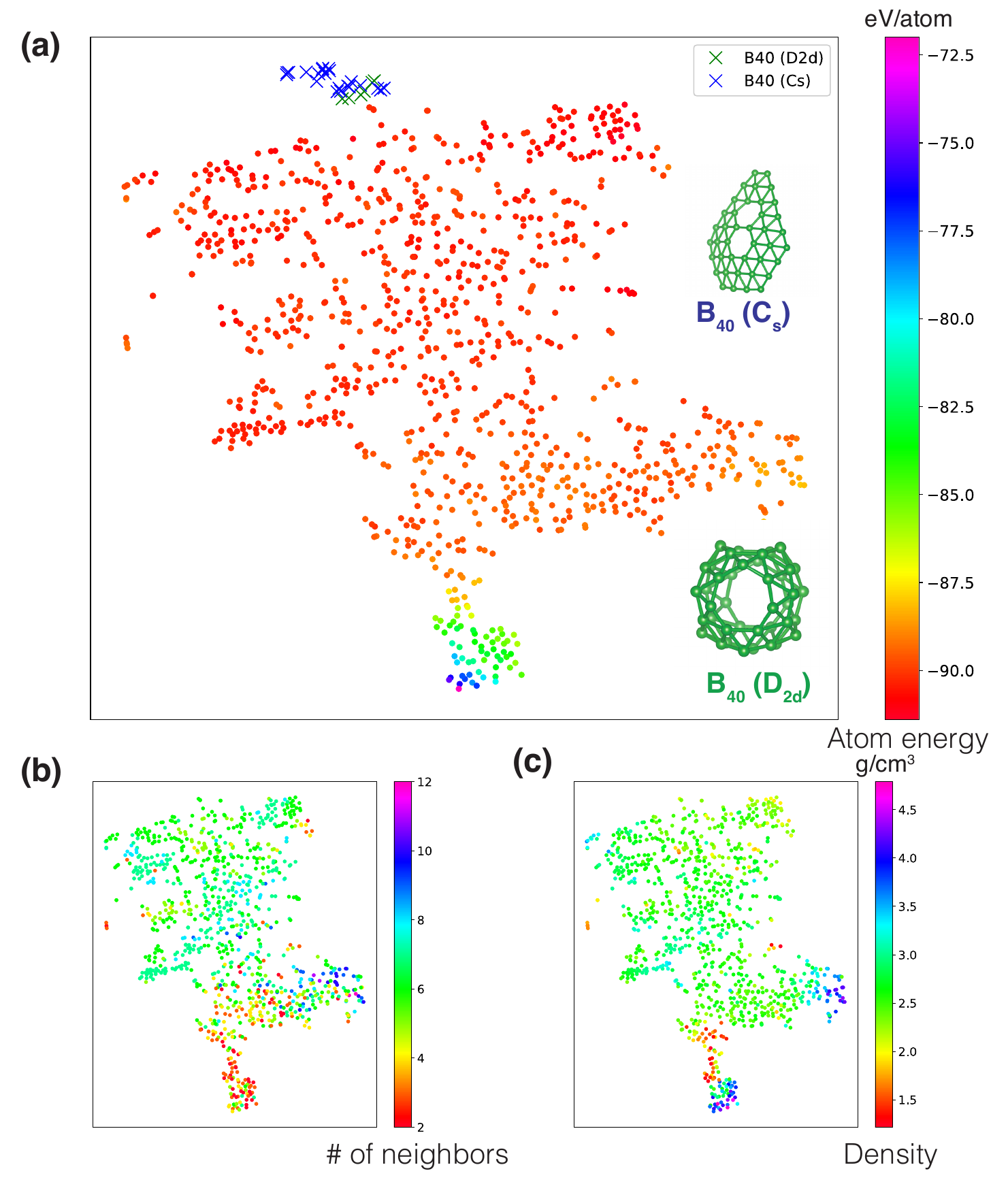}
  \caption{The boron fullerene local environments in the boron structural space. The representation of each distinct local environments in the two \ce{B_40} structures are plotted in the original boron structural space in Fig.~4. }
  \label{fig:si-B-fullerene}
\end{figure}

\begin{figure}[tbh]
  \centering
  \includegraphics[width=\linewidth]{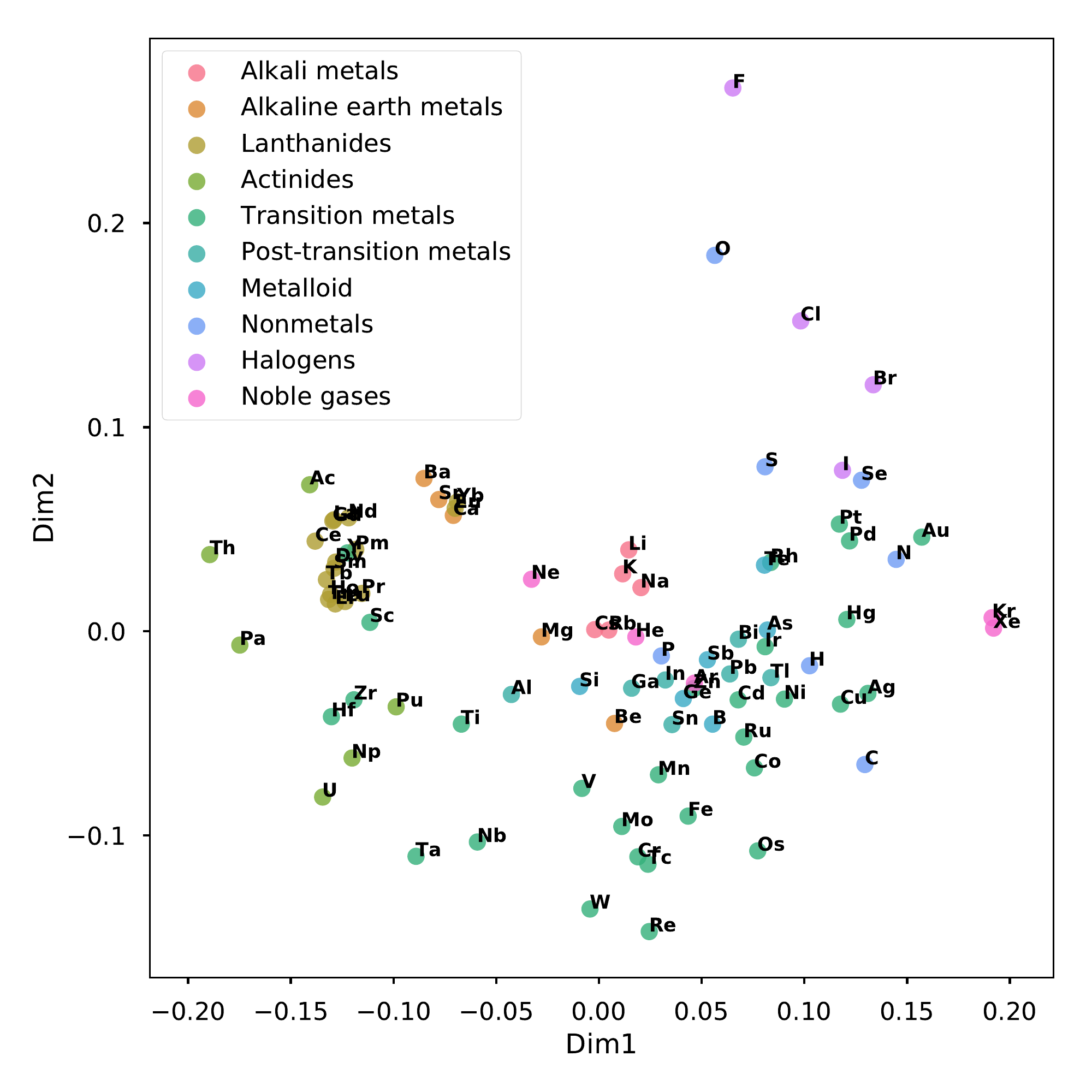}
  \caption{Visualization of the two principal dimensions of the element representations learned from the Materials Project dataset using principal component analysis. }
  \label{fig:si-matproj-pca-vis}
\end{figure}

\begin{figure}[tbh]
  \centering
  \includegraphics[width=\linewidth]{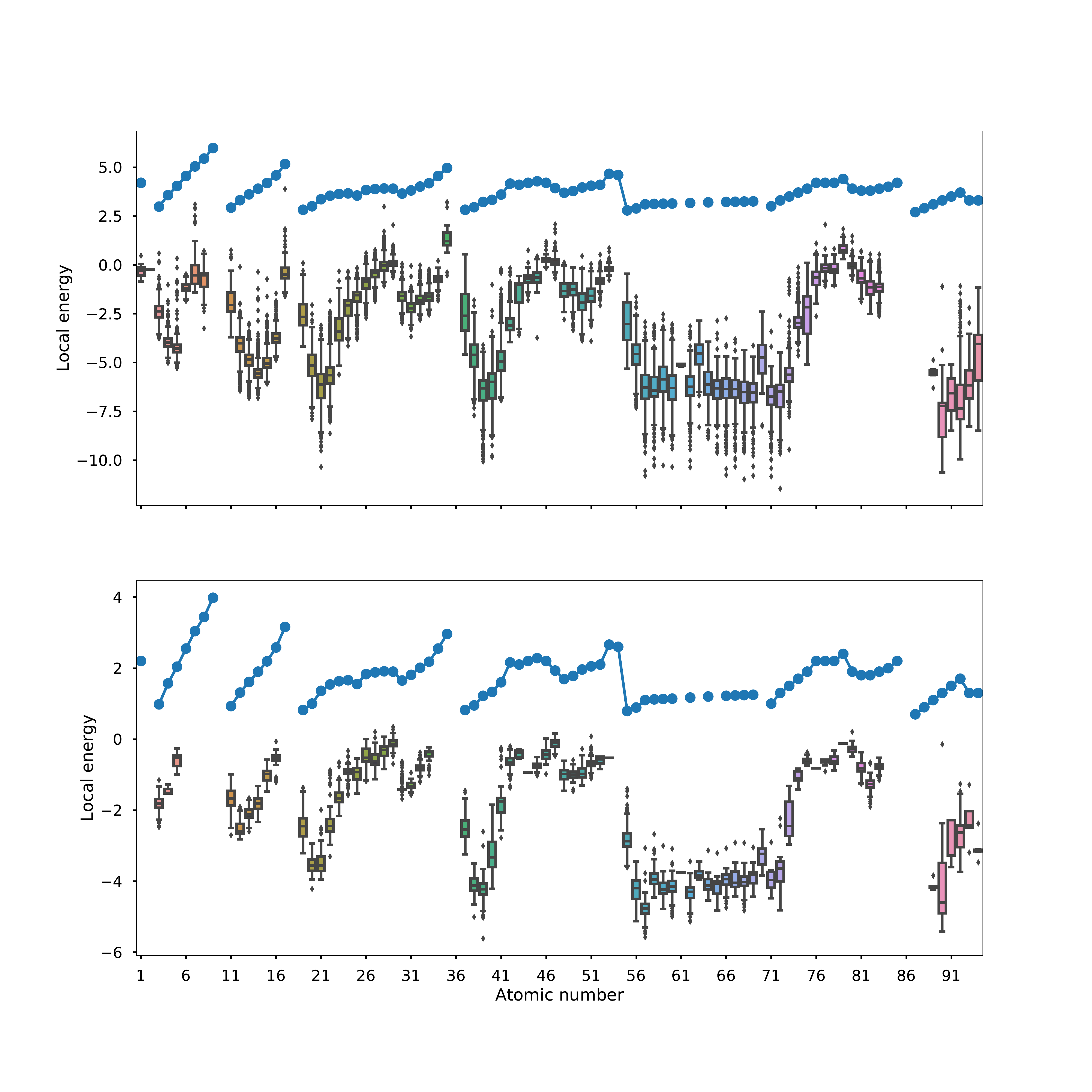}
  \caption{The local energy of oxygen (upper) and sulfur (lower) coordination environments as a function of atomic number. The blue dotted line denotes the electronegativity of each element. }
  \label{fig:si-matproj-pca-vis}
\end{figure}

\end{document}